\documentclass[runningheads]{llncs}
\usepackage{graphicx}
\usepackage{glossaries}
\usepackage{algorithm}
\usepackage{algorithmicx}
\usepackage[noend]{algpseudocode}
\usepackage[normalem]{ulem}
\usepackage{xcolor}
\usepackage{hyperref}

\useunder{\uline}{\ul}{}
\newacronym{dl}{DL}{Deep Learning}
\newacronym{drl}{DRL}{Disentangled Representation Learning}
\newacronym{mri}{MRI}{Magnetic Resonance Imaging}
\newacronym{ae}{AE}{Auto-Encoder}
\newacronym{vae}{VAE}{Variational Auto-Encoder}
\newacronym{kl}{KL}{Kullback-Leibler}
\newacronym{elbo}{ELBO}{Evidence Lower Bound}
\newacronym{pca}{PCA}{Principal Component Analysis}
\newacronym{gan}{GAN}{Generative Adversarial Network}
\newacronym{nf}{NF}{Normalising Flow}
\newacronym{csd}{CSD}{Content-Style Disentanglement}
\newacronym{in}{IN}{Instance Normalisation}
\newacronym{adain}{AdaIN}{Adaptive Instance Normalisation}
\newacronym{film}{FiLM}{Feature-wise Linear Modulation}
\newacronym{spade}{SPADE}{Spatially-Adaptive Denormalisation}
\newacronym{fft}{FFT}{Fast Fourier Transform}
\newacronym{ct}{CT}{Computerised Tomography}
\newacronym{sdnet}{SDNet}{Spatial Decomposition Network}
\newacronym{acdc}{ACDC}{Automatic Cardiac Diagnosis Challenge}
\newacronym{mmwhs}{MM-WHS}{Mulit-Modal Whole Heart Segmentation}
\newacronym{acs}{ACS}{Adversarial Continual Segmenter}
\newacronym{adni}{ADNI}{Alzheimer's Disease Neuroimaging Initiative}
\newacronym{ms}{MS}{Multiple Sclerosis}
\newacronym{wm}{WM}{White Matter}
\newacronym{dnn}{DNN}{Deep Neural Network}
\newacronym{2d}{2D}{2-dimensional}
\newacronym{xai}{XAI}{Explainable AI}
\newacronym{gradcam}{GradCAM}{Gradient-weighted Class Activation Mapping}
\newacronym{cnn}{CNN}{Convolutional Neural Network}
\newacronym{ce}{CE}{Cross-Entropy}
\newacronym{dsc}{DSC}{Dice Score}
\newacronym{ui}{UI}{User Interface}

\begin{document}
\title{Interpretability-guided Data Augmentation for Robust Segmentation in Multi-centre Colonoscopy Data}

\titlerunning{Interpretability-guided Data Augmentation for Multi-centre Data}
% If the paper title is too long for the running head, you can set
% an abbreviated paper title here
%
\author{Valentina Corbetta\inst{1,2}\orcidID{0000-0002-3445-3011} \and
Regina Beets-Tan\inst{1,2}\orcidID{0000-0002-8533-5090} \and
Wilson Silva\inst{1}\orcidID{0000-0002-4080-9328}} 
\authorrunning{V. Corbetta et al.}
% First names are abbreviated in the running head.
% If there are more than two authors, 'et al.' is used.
%
\institute{Department of Radiology, The Netherlands Cancer Institute,
Amsterdam, The Netherlands \\
\email{\{v.corbetta\}@nki.nl}\\ \and
GROW School for Oncology and Developmental Biology,
Maastricht University Medical Center, Maastricht, The Netherlands\\
}
\maketitle              % typeset the header of the contribution

\begin{abstract}
Multi-centre colonoscopy images from various medical centres exhibit distinct complicating factors and overlays that impact the image content, contingent on the specific acquisition centre. Existing Deep Segmentation networks struggle to achieve adequate generalizability in such data sets, and the currently available data augmentation methods do not effectively address these sources of data variability. As a solution, we introduce an innovative data augmentation approach centred on interpretability saliency maps, aimed at enhancing the generalizability of Deep Learning models within the realm of multi-centre colonoscopy image segmentation. The proposed augmentation technique demonstrates increased robustness across different segmentation models and domains. Thorough testing on a publicly available multi-centre dataset for polyp detection demonstrates the effectiveness and versatility of our approach, which is observed both in quantitative and qualitative results. The code is publicly available at: \url{https://github.com/nki-radiology/interpretability_augmentation}

\end{abstract}

\section{Introduction}\label{sec:intro}

The adoption of \gls{dl} techniques has significantly advanced medical image segmentation in recent years~\cite{chen2020deep,liu2021review}. UNet and other U-shaped architectures have been pivotal in this revolution~\cite{liu2020survey}, remaining competitive even with the introduction of newer models.

However, when \gls{dl} models are applied to unseen datasets acquired from different scanners or clinical centers, their performance at inference time declines noticeably~\cite{campello2021multi,prados2017spinal}. This is due to \textit{domain shifts}, caused by variations in data statistics between different clinical centers, resulting from varying patient populations, scanners, and scan settings~\cite{zhang2020generalizing,tao2019deep}. These disparities in patient characteristics and imaging settings can significantly affect the model's ability to generalize effectively~\cite{li2020self,zhang2020generalizing}.

To further integrate \gls{dl} models into clinical practice, it is crucial for them to be robust against these changes and demonstrate a high level of generalizability. The most straightforward approach to address domain shifts is by collecting and annotating as many varied samples as possible. Nevertheless, acquiring and labeling enough data to encompass real-world variation is prohibitively time-consuming and costly. 

\begin{figure}[!t]
\includegraphics[width=\textwidth]{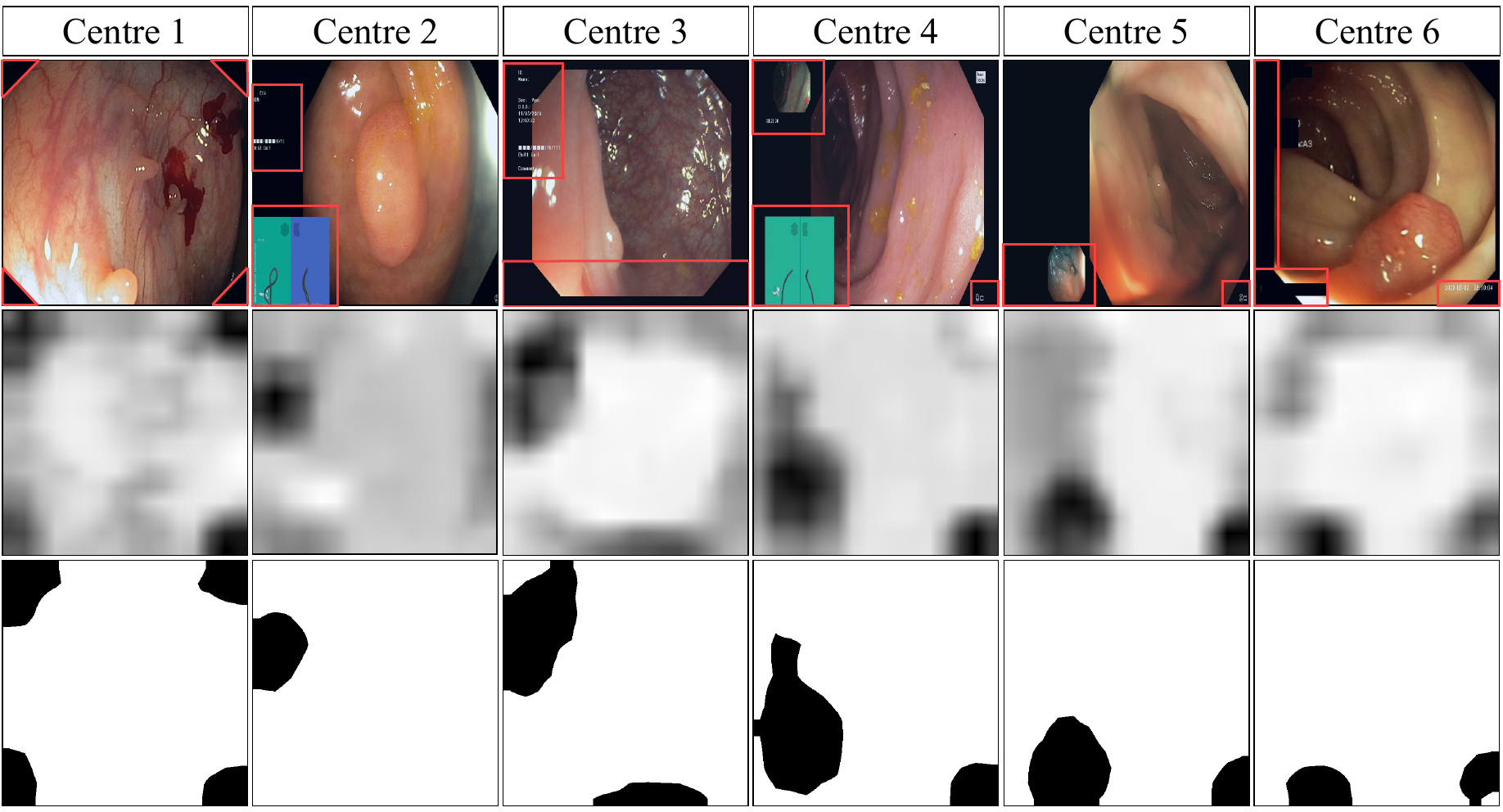}
\caption{Illustration of examples of the "extra"-anatomical content across the different centres in the PolypGen dataset and its impact on the GradCAM visualizations. The first row depicts the original image, the second and third row show the GradCAM visualizations and their binarization, respectively.} \label{fig:centres_difference}
\end{figure}

In recent years, \gls{drl} has emerged as a promising solution to address the aforementioned limitations. This method encodes underlying variation factors into separate latent variables, capturing valuable information relevant to the task at hand. By adopting \gls{drl}, \gls{dl} models gain increased robustness against domain shifts, reducing the need for a large number of meticulously labeled samples~\cite{liu2022learning}. Various \gls{drl} models have been employed for segmentation in the context of multi-centre datasets, yielding state-of-the-art outcomes. One such model is the \gls{sdnet}, which decomposes \gls{2d} medical images into spatial anatomical factors (content) and non-spatial modality factors (style)~\cite{chartsias2019disentangled}. Expanding upon \gls{sdnet}, Jiang et al.~\cite{jiang2020semi} have made additional advancements by further disentangling the pathology factor from the anatomy, particularly when the ground truth mask for anatomy is available. To further improve generalizability, Liu et al.~\cite{liu2021semi} combined \gls{drl} with meta-learning, while Shin et al.~\cite{shin2021unsupervised} have effectively disentangled intensity and non-intensity factors to enable domain adaptation in \gls{ct} images. 

Despite the significant advancements made by \gls{drl} methods in improving model generalizability, it is important to acknowledge that these methods assume that the shift introduced by unseen domains is embedded within the "style" features. However, this assumption does not always hold true, especially in scenarios like videos and images of colonoscopies, or other endoscopy applications. In such cases, various confounding factors affect the content of the images, depending on the domain, from hereon also referred to as \textit{centre}, of acquisition. These factors include image miniaturization, anonymization, and depictions of the instrument's position during image acquisition, as shown in the first row of  Figure~\ref{fig:centres_difference}. As illustrated in Section~\ref{sec:res}, both traditional methods, like UNet and DeepLabV3+~\cite{chen2018encoder}, and \gls{drl} models like \gls{sdnet} encounter challenges in generalizing to domains heavily characterized by this additional content that is unrelated to the anatomy.

To address these limitations, we propose an innovative data augmentation strategy based on interpretability techniques. Interpretability techniques have already been successfully applied to improve \gls{dl} models' performance in medical image analysis tasks: Silva et al.~\cite{silva2020interpretability} exploited interpretability methods to improve medical image retrieval in the radiological workflow; in~\cite{9744030}, \glspl{gradcam} are used to improve generalized zero shot learning for medical image classification. Firstly, we pretrain a classifier network to identify the respective centres to which the images belong, using the same training set that will later be employed to train the segmentation module. Thus, we ensure a fair assessment of the segmentation model. During the training phase of the segmentation network, we employ the pre-trained classifier to generate visual explanations for the input batch. In this work, we use \gls{gradcam}~\cite{selvaraju2017grad} to produce the visual explanations, a widely adopted technique for visualizing and interpreting the decision-making process of \glspl{cnn} in a wide variety of computer vision tasks. By leveraging gradients of the predicted class, \gls{gradcam} assigns importance weights to different spatial locations within the last convolutional layer. This allows us to identify the regions in the input image that significantly contribute to the model's decision-making process. Figure~\ref{fig:centres_difference}, specifically the second row, showcases examples of the generated \gls{gradcam} visualizations for each centre. Notably, the classifier predominantly focuses on areas where the "extra"-anatomical content resides (indicated by darker regions). We binarize the generated \glspl{gradcam}, as depicted in the third row of Figure~\ref{fig:centres_difference}, and multiply them with a probability $p$ with the input to the segmentation network. Thus, this approach randomly blocks out the additional information, enabling the segmentation network to place greater emphasis on the anatomical regions.

\begin{figure}[!t]
\includegraphics[width=\textwidth]{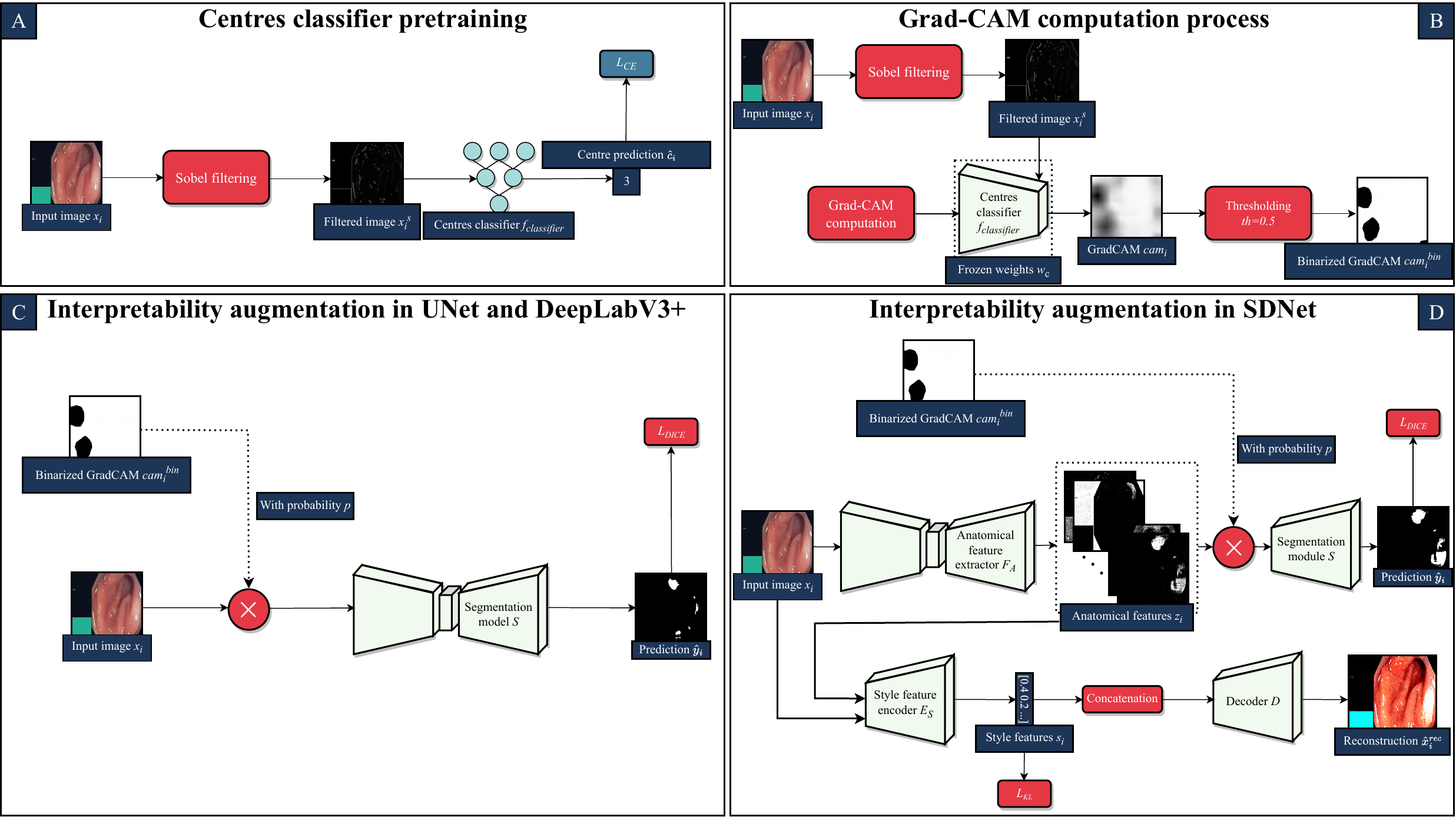}
\caption{Illustration of the proposed method. In \textbf{A}, the pretraining of the classifier is shown. \textbf{B} depicts the process of computing GradCAM visualizations. As for \textbf{C} and \textbf{D}, they demonstrate the incorporation of the interpretability-guided augmentation in the training phase of UNet and DeepLabV3+, and SDNet, respectively.} \label{fig:method}
\end{figure}

The key contributions of our research can be summarized as follows:
\begin{itemize}
    \item We introduce a novel data augmentation technique based on interpretability techniques. By incorporating visual explanations, we develop a robust technique that can be readily applied to different and multiple domains. This stands in contrast to standard augmentation techniques, which are limited in their ability to capture the variability of non-synthetic data~\cite{zhang2020generalizing}.

    \item We apply and adapt the proposed methodology to both two conventional baseline models, UNet and DeepLabV3+ with ResNet101 as backbone, and a \gls{drl} model, specifically \gls{sdnet}. This showcases the versatility of our augmentation strategy across different architectures.

    \item We conduct thorough testing of our method using an open-source multi-centre dataset, PolypGen~\cite{ali2023multi}, to demonstrate the robustness of our technique and its effectiveness in diverse domains.
\end{itemize}

\section{Methodology}

The proposed methodology is illustrated in Figure~\ref{fig:method}, outlining the pretraining of the classifier module, the generation process of the \gls{gradcam} visualizations and their integration into the UNet, DeepLabV3+ and \gls{sdnet} models.

\subsection{Pretraining of Centres Classifier}

For the classifier backbone, we employ a ResNet50 architecture~\cite{he2016deep} with pretraining on ImageNet~\cite{deng2009imagenet}. The module is trained on the same training set used for subsequent training of the segmentation networks. Given an input image $x_i$, we initially apply Sobel filtering~\cite{996} to emphasize the edges characterizing the extra-anatomical content and discard most of the anatomy. The resulting filtered image $x_i^s$ is then passed through the ResNet50 to predict the original centre to which the image belongs. The classification network is trained using the \gls{ce} Loss between the predicted $\hat{c}_i$ and the original centre label $c_i$.

\subsection{GradCAM Visualizations Generation Process}

\begin{algorithm}[!t]
  \caption{Gradient-weighted Class Activation Mapping (GradCAM)}
  \begin{algorithmic}[1]
    \Procedure{GradCAM}{$\mathbf{x_i^s}$, $\mathbf{w_c}$}
      \State $\mathbf{g} \gets$ Compute the gradient of the target class score $c$ w.r.t the feature map $a$ of the last convolutional layer $l_c$
      \State $\mathbf{i_w} \gets$ Compute the importance vector of each feature channel by applying global average pooling to $\mathbf{g}$
      \State $\mathbf{cam_i} \gets$ Compute the class activation map by combining the importance vector $\mathbf{i_w}$ with the corresponding feature map $a$
      \State $\mathbf{cam_i} \gets \text{ReLU}(\mathbf{cam_i})$ \Comment{Apply ReLU activation to remove negative values}
      \State $\mathbf{cam_i} \gets \text{normalize}(\mathbf{cam_i})$ \Comment{Normalize the class activation map}
      \State $\mathbf{cam_i} \gets \text{upsample}(\mathbf{cam_i})$ \Comment{Upasample to match size of $\mathbf{x_i^s}$}
      \State \textbf{return} $\mathbf{cam_i}$ \Comment{Return the final GradCAM visualization}
    \EndProcedure
  \end{algorithmic}
  \label{alg:gradcam}
\end{algorithm}

During the training process of the segmentation network, the pretrained weights $w_c$ of the centres classifier are loaded and kept frozen. The classifier is then used to perform inference on the Sobel-filtered input image $x_i^s$. Subsequently, the \gls{gradcam} visualization $cam_i$ is generated following the steps outlined in Algorithm~\ref{alg:gradcam}. The \gls{gradcam} visualizations provide a coarse representation of the areas in the input image $x_i^s$ that the classifier focused on to make its prediction. To ensure that we do not inadvertently block useful content for the downstream segmentation task, we binarize $cam_i$ using a threshold $th=0.5$ to obtain $cam_i^{bin}$. This ensures that the augmentation procedure described in the subsequent paragraphs masks only the most relevant "extra"-anatomical content.

\subsection{Interpretability-guided Data Augmentation}

We will now provide a detailed explanation of how the \gls{gradcam} visualizations are utilized as an augmentation technique to enhance the robustness and generalizability of segmentation models.

\subsubsection{UNet and DeepLabV3+} 
The integration of the \gls{gradcam} visualizations into the UNet and DeepLabV3+ training process is straightforward. With a probability $p$, we multiply the input image $x_i$ by the corresponding \gls{gradcam} visualization $cam_i^{bin}$.
We introduce a probability $p$ for the multiplication step to mitigate the risk of covering important areas for the downstream task. Figure~\ref{fig:example_aug} provides two examples of augmented samples in the UNet and DeepLabV3+ training process. The models are trained by computing the Dice Loss between the predicted segmentation mask $\hat{y}_i$ and the ground truth $y_i$.

\subsubsection{SDNet} 

\begin{figure}[!t]
\includegraphics[width=\textwidth]{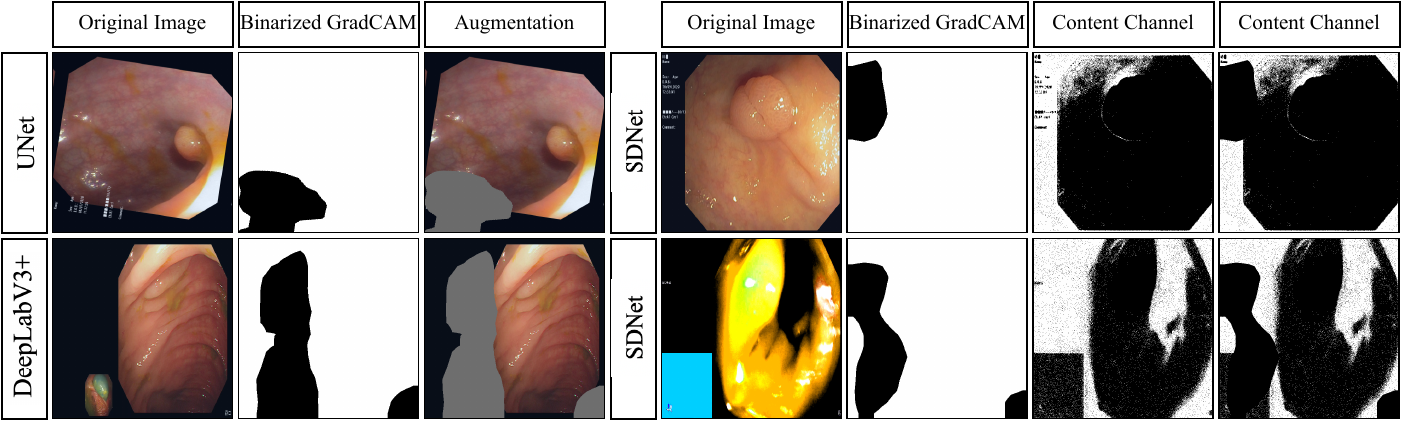}
\caption{Examples of augmented samples with our interpretability-guided augmentation, for UNet, DeepLabV3+ and SDNet.} \label{fig:example_aug}
\end{figure}

To gain a better understanding of how interpretability-guided augmentation is integrated into the \gls{sdnet}, it is necessary to provide a brief overview of the model's structure. Initially, the anatomy encoder $F_{anatomy}$ encodes input image $x_i$ into a multi-channel spatial representation, the anatomical features $z_i$. It is important to note that $z_i$ has shape $N \times H \times W$, where $N$ represents a fixed number of channels (e.g. 8), while $H$ and $W$ correspond to the height and width of the original image, respectively. The modality encoder $E_{s}$ uses factor $z_i$ along with the input image $x_i$ to produce the latent vector $s_i$, representing the style features. These two representations, $s_i$ and $z_i$, are combined to reconstruct the input image through the decoder network $D$. The anatomical representation $z_i$ is then fed into the segmentation network $S$ to produce the segmentation mask $\hat{y}_i$. For a more detailed explanation of the \gls{sdnet} and its associated losses, we refer the reader to the schematic in Figure~\ref{fig:method}(D) and to the original paper~\cite{chartsias2019disentangled}. 
To perform the interpretability-guided augmentation, we multiply the multi-channel anatomical representation $z_i$ with the binary \gls{gradcam} visualization $cam_i^{bin}$ using a probability $p$ before feeding it as input to the segmentation module $S$. Figure~\ref{fig:example_aug} provides two examples of augmented samples in the \gls{sdnet} training process, in particular we report the effect on only one of the $N$ channels of $z_i$ for synthesis purposes. We made the decision to apply the augmentation on the anatomy representation $z_i$ and not on the input image $x_i$. This approach is intended to mimic the process used in the UNet and DeepLabv3+, where we directly manipulate the input to the module dedicated to the downstream segmentation task, while keeping the rest of the \gls{sdnet} architecture intact. 
\section{Results}\label{sec:res}

\begin{table}[t]
\caption{The comparison results of the proposed method with the corresponding baseline models. The best result concerning the specific baseline considered is highlighted in \textbf{bold}. In the Dice column, the value of probability $p$ that produced the best result in terms of Dice score in the interpretability-guided augmentation is indicated in parentheses.}
\resizebox{\textwidth}{!}{%
\begin{tabular}{c|ccc|ccc|llllll}
\cline{2-7}
\multicolumn{1}{l|}{}                       & \multicolumn{1}{c|}{\textbf{Dice}}   & \multicolumn{1}{c|}{\textbf{Recall}} & \textbf{Accuracy} & \multicolumn{1}{c|}{\textbf{Dice}}          & \multicolumn{1}{c|}{\textbf{Recall}} & \textbf{Accuracy} &  &  &  &  &  &  \\ \cline{1-7}
\multicolumn{1}{|c|}{\textbf{Out-dist set}} & \multicolumn{3}{c|}{\textbf{UNet}}                                                              & \multicolumn{3}{c|}{\textbf{UNet interpretability augmentation}}                                       &  &  &  &  &  &  \\ \cline{1-7}
\multicolumn{1}{|c|}{centre 1}              & \multicolumn{1}{c|}{0.7257}          & \multicolumn{1}{c|}{0.7725}          & 0.9589            & \multicolumn{1}{c|}{\textbf{0.7353 (60\%)}} & \multicolumn{1}{c|}{0.7716}          & 0.9587            &  &  &  &  &  &  \\ \cline{1-7}
\multicolumn{1}{|c|}{centre 2}              & \multicolumn{1}{c|}{0.5762}          & \multicolumn{1}{c|}{\textbf{0.6454}} & 0,9440             & \multicolumn{1}{c|}{\textbf{0.6062 (60\%)}} & \multicolumn{1}{c|}{0.5948}          & \textbf{0.9530}    &  &  &  &  &  &  \\ \cline{1-7}
\multicolumn{1}{|c|}{centre 3}              & \multicolumn{1}{c|}{0.7054}          & \multicolumn{1}{c|}{0.6724}          & 0.9619            & \multicolumn{1}{c|}{\textbf{0.7470 (40\%)}}  & \multicolumn{1}{c|}{\textbf{0.6966}} & \textbf{0.9658}   &  &  &  &  &  &  \\ \cline{1-7}
\multicolumn{1}{|c|}{centre 4}              & \multicolumn{1}{c|}{0.4223}          & \multicolumn{1}{c|}{\textbf{0.3804}} & 0.9594            & \multicolumn{1}{c|}{\textbf{0.4519 (40\%)}} & \multicolumn{1}{c|}{0.3757}          & \textbf{0.9607}   &  &  &  &  &  &  \\ \cline{1-7}
\multicolumn{1}{|c|}{centre 5}              & \multicolumn{1}{c|}{0.4725}          & \multicolumn{1}{c|}{0.4934}          & 0.9581            & \multicolumn{1}{c|}{\textbf{0.4893 (40\%)}} & \multicolumn{1}{c|}{\textbf{0.5369}} & 0.9583            &  &  &  &  &  &  \\ \cline{1-7}
\multicolumn{1}{|c|}{centre 6}              & \multicolumn{1}{c|}{0.6423}          & \multicolumn{1}{c|}{0.5597}          & 0.9593            & \multicolumn{1}{c|}{\textbf{0.6574 (60\%)}} & \multicolumn{1}{c|}{\textbf{0.6132}} & \textbf{0.9634}   &  &  &  &  &  &  \\ \cline{1-7}
\multicolumn{1}{l|}{}                       & \multicolumn{1}{c|}{\textbf{Dice}}   & \multicolumn{1}{c|}{\textbf{Recall}} & \textbf{Accuracy} & \multicolumn{1}{c|}{\textbf{Dice}}          & \multicolumn{1}{c|}{\textbf{Recall}} & \textbf{Accuracy} &  &  &  &  &  &  \\ \cline{1-7}
\multicolumn{1}{|c|}{\textbf{Out-dist set}} & \multicolumn{3}{c|}{\textbf{Deeplabv3+}}                                                        & \multicolumn{3}{c|}{\textbf{Deeplabv3+ interpretability augmentation}}                                 &  &  &  &  &  &  \\ \cline{1-7}
\multicolumn{1}{|c|}{centre 1}              & \multicolumn{1}{c|}{0.6003}          & \multicolumn{1}{c|}{\textbf{0.6284}} & 0.9482            & \multicolumn{1}{c|}{\textbf{0.6155 (60\%)}} & \multicolumn{1}{c|}{0.6195}          & \textbf{0.9545}   &  &  &  &  &  &  \\ \cline{1-7}
\multicolumn{1}{|c|}{centre 2}              & \multicolumn{1}{c|}{0.5498}          & \multicolumn{1}{c|}{\textbf{0.5358}} & 0.9398            & \multicolumn{1}{c|}{\textbf{0.5679 (60\%)}} & \multicolumn{1}{c|}{0.5159}          & \textbf{0.9496}   &  &  &  &  &  &  \\ \cline{1-7}
\multicolumn{1}{|c|}{centre 3}              & \multicolumn{1}{c|}{0.5696}          & \multicolumn{1}{c|}{0.5407}          & \textbf{0.9585}   & \multicolumn{1}{c|}{\textbf{0.6314 (40\%)}} & \multicolumn{1}{c|}{\textbf{0.6745}} & 0.9561            &  &  &  &  &  &  \\ \cline{1-7}
\multicolumn{1}{|c|}{centre 4}              & \multicolumn{1}{c|}{0.3424}          & \multicolumn{1}{c|}{\textbf{0.2471}} & 0.9483            & \multicolumn{1}{c|}{\textbf{0.3592 (50\%)}} & \multicolumn{1}{c|}{0.2166}          & \textbf{0.9493}   &  &  &  &  &  &  \\ \cline{1-7}
\multicolumn{1}{|c|}{centre 5}              & \multicolumn{1}{c|}{0.3867}          & \multicolumn{1}{c|}{0.4171}          & \textbf{0.9543}   & \multicolumn{1}{c|}{\textbf{0.4046 (60\%)}} & \multicolumn{1}{c|}{\textbf{0.4479}} & 0.9533            &  &  &  &  &  &  \\ \cline{1-7}
\multicolumn{1}{|c|}{centre 6}              & \multicolumn{1}{c|}{0.6268}          & \multicolumn{1}{c|}{0.5932}          & 0.9631            & \multicolumn{1}{c|}{\textbf{0.6342 (40\%)}} & \multicolumn{1}{c|}{\textbf{0.6126}} & \textbf{0.9658}   &  &  &  &  &  &  \\ \cline{1-7}
\multicolumn{1}{l|}{}                       & \multicolumn{1}{c|}{\textbf{Dice}}   & \multicolumn{1}{c|}{\textbf{Recall}} & \textbf{Accuracy} & \multicolumn{1}{c|}{\textbf{Dice}}          & \multicolumn{1}{c|}{\textbf{Recall}} & \textbf{Accuracy} &  &  &  &  &  &  \\ \cline{1-7}
\multicolumn{1}{|c|}{\textbf{Out-dist set}} & \multicolumn{3}{c|}{\textbf{SDNet}}                                                             & \multicolumn{3}{c|}{\textbf{SDNet interpretability augmentation}}                                      &  &  &  &  &  &  \\ \cline{1-7}
\multicolumn{1}{|c|}{centre 1}              & \multicolumn{1}{c|}{0.7130}           & \multicolumn{1}{c|}{0.7583}          & 0.9551            & \multicolumn{1}{c|}{\textbf{0.7226 (40\%)}} & \multicolumn{1}{c|}{\textbf{0.7726}} & \textbf{0.9575}   &  &  &  &  &  &  \\ \cline{1-7}
\multicolumn{1}{|c|}{centre 2}              & \multicolumn{1}{c|}{0.5489}          & \multicolumn{1}{c|}{\textbf{0.5794}} & 0.9328            & \multicolumn{1}{c|}{\textbf{0.5579 (50\%)}} & \multicolumn{1}{c|}{0.5482}          & \textbf{0.9464}   &  &  &  &  &  &  \\ \cline{1-7}
\multicolumn{1}{|c|}{centre 3}              & \multicolumn{1}{c|}{0.7151}          & \multicolumn{1}{c|}{0.7082}          & \textbf{0.9620}    & \multicolumn{1}{c|}{\textbf{0.7208 (40\%)}} & \multicolumn{1}{c|}{\textbf{0.7245}} & 0.9603            &  &  &  &  &  &  \\ \cline{1-7}
\multicolumn{1}{|c|}{centre 4}              & \multicolumn{1}{c|}{\textbf{0.3981}} & \multicolumn{1}{c|}{\textbf{0.3254}} & \textbf{0.9591}   & \multicolumn{1}{c|}{0.3841 (50\%)}          & \multicolumn{1}{c|}{0.3360}           & 0.9557            &  &  &  &  &  &  \\ \cline{1-7}
\multicolumn{1}{|c|}{centre 5}              & \multicolumn{1}{c|}{0.4312}          & \multicolumn{1}{c|}{0.4293}          & 0.9587            & \multicolumn{1}{c|}{\textbf{0.4546 (40\%)}} & \multicolumn{1}{c|}{\textbf{0.4722}} & 0.9588            &  &  &  &  &  &  \\ \cline{1-7}
\multicolumn{1}{|c|}{centre 6}              & \multicolumn{1}{c|}{0.6398}          & \multicolumn{1}{c|}{\textbf{0.6195}} & 0.9610             & \multicolumn{1}{c|}{\textbf{0.6626 (60\%)}} & \multicolumn{1}{c|}{0.5977}          & \textbf{0.9621}   &  &  &  &  &  &  \\ \cline{1-7}
\end{tabular}%
}
\label{tab:results}
\end{table}

\subsubsection{Dataset and Implementation Details}

To evaluate the proposed methodology, we utilized the publicly available PolypGen dataset, which comprises colonoscopy data collected from 6 different centres, encompassing diverse patient populations. Our analysis focused on single frame samples, resulting in a total of 1537 images (Centre 1: 256, Centre 2: 301, Centre 3: 457, Centre 4: 227, Centre 5: 208, Centre 6: 88). Notably, as illustrated in Figure~\ref{fig:centres_difference}, the centres exhibit substantial variability in image content, both within and across centres. To assess the generalizability of the models, we conducted 6 distinct experiments for each tested model: we trained and validated the models using 5 centres while reserving one centre as an out-of-distribution test set. Throughout the experiments, a patient-level split was applied for the in-distribution frames, with 80\% of patients allocated for training, 10\% for validation, and 10\% for the in-distribution test set.

Standard preprocessing techniques were applied to the images, including resizing them to a dimension of $256 \times 256$ and normalizing the pixel values. Furthermore, additional augmentations were performed exclusively on the training set. These augmentations comprised rotation, horizontal and vertical flips (with a 50\% probability), which were also applied to the ground truth and binary \gls{gradcam} visualizations. Additionally, colour jitter was applied to the images with a 30\% probability. The models were trained for 300 epochs, except for the centres classifier which was trained just for 10 epochs, on an NVIDIA RTX\textsuperscript{\texttrademark} A6000 GPU, utilizing a batch size of 4 and a learning rate of $10^{-5}$.

\subsubsection{Results and Discussion}

\begin{figure}[!t]
\includegraphics[width=\textwidth]{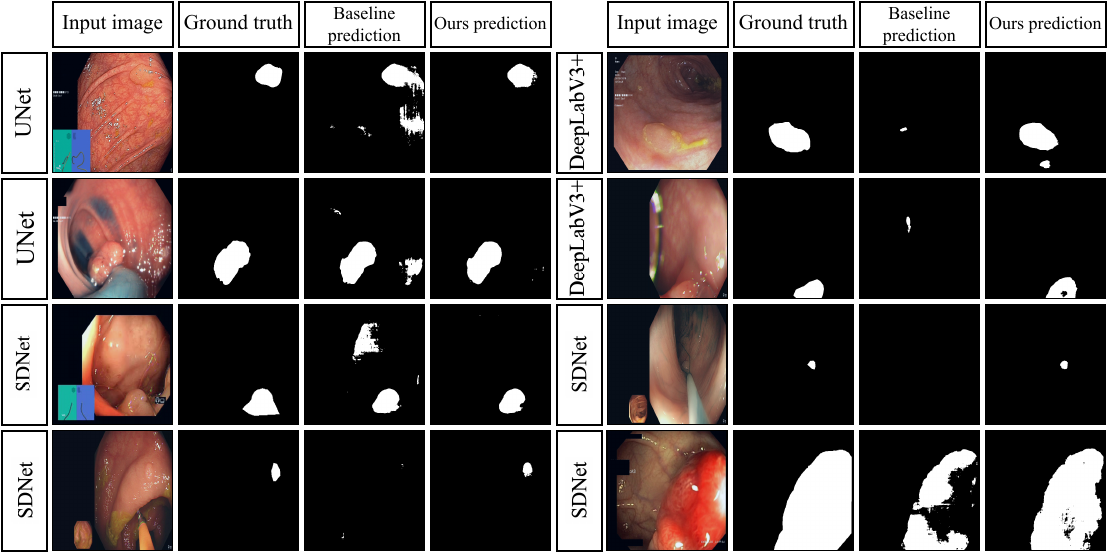}
\caption{Examples of qualitative results for the baselines and the interpretability-guided augmentation approach.} \label{fig:qualitative_res}
\end{figure}

The experimental results are presented in Table~\ref{tab:results}, where we evaluate segmentation outcomes using as metrics the \gls{dsc}, recall, and accuracy. To determine the best-performing augmented models based on \gls{dsc}, we conduct a parameter study on the probability value $p$, as elaborated in the following paragraph, and subsequently compare them with the corresponding baseline models. Concerning the \gls{dsc}, the models trained with interpretability-guided augmentation demonstrate superior performance in nearly all experiments. Particularly noteworthy is the substantial improvement in Centre 3 for the UNet, with an increase of 4.16\%. Additionally, the \gls{sdnet} exhibits an increment of 2.57\% in Centre 5, and for DeepLabV3+, there is a significant 6.18\% increase in Centre 3. The results also consistently demonstrate improvements in terms of accuracy and recall in nearly all experiments. Figure~\ref{fig:qualitative_res} presents several qualitative examples of the segmentation results. Notably, the masks obtained using our proposed methodology exhibit reduced noise levels, and our approach demonstrates greater performance in detecting smaller polyps.

\subsubsection{Parameters Study}

We delve deeper into the effectiveness of our proposed technique, conducting a detailed study of the parameters for the all the analyzed architectures. Our augmentation method was employed with probabilities of 40\%, 50\% and 60\%. The results of this study are displayed in the Supplementary Material. The study proves that the fine-tuning of the probability value $p$ within our augmentation approach plays a pivotal role in enhancing the models' generalizability. Indeed, when the suitable probability $p$ is applied, the augmented architectures surpass the performance of the baseline models in nearly all the  tests.

\section{Conclusion}

We have introduced an interpretability-guided augmentation technique aimed at improving the generalizability of \gls{dl} models to unseen domains in colonoscopy segmentation tasks. We have demonstrated the strength and reliability of this proposed technique by successfully adapting it to three distinct architectures: UNet, \gls{sdnet} and DeeplabV3+. Our future work will involve testing the proposed augmentation in low-data and semi-supervised settings, where \gls{drl} models, and in particular \gls{sdnet}, significantly outperform conventional models. Moreover, we plan to investigate the adaptability of our methodology to other multi-centre endoscopy images, such as cystoscopy or laparoscopy, as they exhibit the same instrumentation-specific and \gls{ui} overlays as colonoscopy data. Consequently, our method could improve models' generalizability to such data. 

\subsubsection*{Acknowledgements}
\footnotesize
Research at the Netherlands Cancer Institute is supported by grants from the Dutch Cancer Society, the Dutch Ministry of Health, Welfare and Sport and private sectors.

% ---- Bibliography ----
%
% BibTeX users should specify bibliography style 'splncs04'.
% References will then be sorted and formatted in the correct style.
%
\bibliographystyle{splncs04}
\bibliography{mybibliography}

\begin{thebibliography}{10}
\providecommand{\url}[1]{\texttt{#1}}
\providecommand{\urlprefix}{URL }
\providecommand{\doi}[1]{https://doi.org/#1}

\bibitem{ali2023multi}
Ali, S., Jha, D., Ghatwary, N., Realdon, S., Cannizzaro, R., Salem, O.E.,
  Lamarque, D., Daul, C., Riegler, M.A., Anonsen, K.V., et~al.: A multi-centre
  polyp detection and segmentation dataset for generalisability assessment.
  Scientific Data  \textbf{10}(1), ~75 (2023)

\bibitem{campello2021multi}
Campello, V.M., Gkontra, P., Izquierdo, C., Martin-Isla, C., Sojoudi, A., Full,
  P.M., Maier-Hein, K., Zhang, Y., He, Z., Ma, J., et~al.: Multi-centre,
  multi-vendor and multi-disease cardiac segmentation: the m\&ms challenge.
  IEEE Transactions on Medical Imaging  \textbf{40}(12),  3543--3554 (2021)

\bibitem{chartsias2019disentangled}
Chartsias, A., Joyce, T., Papanastasiou, G., Semple, S., Williams, M., Newby,
  D.E., Dharmakumar, R., Tsaftaris, S.A.: Disentangled representation learning
  in cardiac image analysis. Medical image analysis  \textbf{58},  101535
  (2019)

\bibitem{chen2020deep}
Chen, C., Qin, C., Qiu, H., Tarroni, G., Duan, J., Bai, W., Rueckert, D.: Deep
  learning for cardiac image segmentation: a review. Frontiers in
  Cardiovascular Medicine  \textbf{7}, ~25 (2020)

\bibitem{chen2018encoder}
Chen, L.C., Zhu, Y., Papandreou, G., Schroff, F., Adam, H.: Encoder-decoder
  with atrous separable convolution for semantic image segmentation. In:
  Proceedings of the European conference on computer vision (ECCV). pp.
  801--818 (2018)

\bibitem{deng2009imagenet}
Deng, J., Dong, W., Socher, R., Li, L.J., Li, K., Fei-Fei, L.: Imagenet: A
  large-scale hierarchical image database. In: 2009 IEEE conference on computer
  vision and pattern recognition. pp. 248--255. Ieee (2009)

\bibitem{he2016deep}
He, K., Zhang, X., Ren, S., Sun, J.: Deep residual learning for image
  recognition. In: Proceedings of the IEEE conference on computer vision and
  pattern recognition. pp. 770--778 (2016)

\bibitem{jiang2020semi}
Jiang, H., Chartsias, A., Zhang, X., Papanastasiou, G., Semple, S., Dweck, M.,
  Semple, D., Dharmakumar, R., Tsaftaris, S.A.: Semi-supervised pathology
  segmentation with disentangled representations. In: Domain Adaptation and
  Representation Transfer, and Distributed and Collaborative Learning: Second
  MICCAI Workshop, DART 2020, and First MICCAI Workshop, DCL 2020, Held in
  Conjunction with MICCAI 2020, Lima, Peru, October 4--8, 2020, Proceedings 2.
  pp. 62--72. Springer (2020)

\bibitem{996}
Kanopoulos, N., Vasanthavada, N., Baker, R.: Design of an image edge detection
  filter using the sobel operator. IEEE Journal of Solid-State Circuits
  \textbf{23}(2),  358--367 (1988). \doi{10.1109/4.996}

\bibitem{li2020self}
Li, Y., Chen, J., Xie, X., Ma, K., Zheng, Y.: Self-loop uncertainty: A novel
  pseudo-label for semi-supervised medical image segmentation. In:
  International Conference on Medical Image Computing and Computer-Assisted
  Intervention. pp. 614--623. Springer (2020)

\bibitem{liu2020survey}
Liu, L., Cheng, J., Quan, Q., Wu, F.X., Wang, Y.P., Wang, J.: A survey on
  u-shaped networks in medical image segmentations. Neurocomputing
  \textbf{409},  244--258 (2020)

\bibitem{liu2021review}
Liu, X., Song, L., Liu, S., Zhang, Y.: A review of deep-learning-based medical
  image segmentation methods. Sustainability  \textbf{13}(3), ~1224 (2021)

\bibitem{liu2022learning}
Liu, X., Sanchez, P., Thermos, S., O’Neil, A.Q., Tsaftaris, S.A.: Learning
  disentangled representations in the imaging domain. Medical Image Analysis
  \textbf{80},  102516 (2022)

\bibitem{liu2021semi}
Liu, X., Thermos, S., O’Neil, A., Tsaftaris, S.A.: Semi-supervised
  meta-learning with disentanglement for domain-generalised medical image
  segmentation. In: Medical Image Computing and Computer Assisted
  Intervention--MICCAI 2021: 24th International Conference, Strasbourg, France,
  September 27--October 1, 2021, Proceedings, Part II 24. pp. 307--317.
  Springer (2021)

\bibitem{9744030}
Mahapatra, D., Ge, Z., Reyes, M.: Self-supervised generalized zero shot
  learning for medical image classification using novel interpretable saliency
  maps. IEEE Transactions on Medical Imaging  \textbf{41}(9),  2443--2456
  (2022). \doi{10.1109/TMI.2022.3163232}

\bibitem{prados2017spinal}
Prados, F., Ashburner, J., Blaiotta, C., Brosch, T., Carballido-Gamio, J.,
  Cardoso, M.J., Conrad, B.N., Datta, E., D{\'a}vid, G., De~Leener, B., et~al.:
  Spinal cord grey matter segmentation challenge. Neuroimage  \textbf{152},
  312--329 (2017)

\bibitem{selvaraju2017grad}
Selvaraju, R.R., Cogswell, M., Das, A., Vedantam, R., Parikh, D., Batra, D.:
  Grad-cam: Visual explanations from deep networks via gradient-based
  localization. In: Proceedings of the IEEE international conference on
  computer vision. pp. 618--626 (2017)

\bibitem{shin2021unsupervised}
Shin, S.Y., Lee, S., Summers, R.M.: Unsupervised domain adaptation for small
  bowel segmentation using disentangled representation. In: Medical Image
  Computing and Computer Assisted Intervention--MICCAI 2021: 24th International
  Conference, Strasbourg, France, September 27--October 1, 2021, Proceedings,
  Part III 24. pp. 282--292. Springer (2021)

\bibitem{silva2020interpretability}
Silva, W., Poellinger, A., Cardoso, J.S., Reyes, M.: Interpretability-guided
  content-based medical image retrieval. In: Medical Image Computing and
  Computer Assisted Intervention--MICCAI 2020: 23rd International Conference,
  Lima, Peru, October 4--8, 2020, Proceedings, Part I 23. pp. 305--314.
  Springer (2020)

\bibitem{tao2019deep}
Tao, Q., Yan, W., Wang, Y., Paiman, E.H., Shamonin, D.P., Garg, P., Plein, S.,
  Huang, L., Xia, L., Sramko, M., et~al.: Deep learning--based method for fully
  automatic quantification of left ventricle function from cine mr images: a
  multivendor, multicenter study. Radiology  \textbf{290}(1),  81--88 (2019)

\bibitem{zhang2020generalizing}
Zhang, L., Wang, X., Yang, D., Sanford, T., Harmon, S., Turkbey, B., Wood,
  B.J., Roth, H., Myronenko, A., Xu, D., et~al.: Generalizing deep learning for
  medical image segmentation to unseen domains via deep stacked transformation.
  IEEE transactions on medical imaging  \textbf{39}(7),  2531--2540 (2020)

\end{thebibliography}

\end{document}